\documentclass[twocolumn,aps,amssymb,prl]{revtex4}
\usepackage{amssymb}
\usepackage{graphicx}
\usepackage{epsfig,amsmath}
\newcommand{\eavg}[1]{\langle{#1}\rangle}

\begin{document}

\title{Shot Noise with Interaction Effects in Single Walled Carbon Nanotubes}

\author{F.~Wu$^1$, P. Queipo$^{2}$, T.~Tsuneta$^{1}$,
T.~H.~Wang$^{3}$, E.~Kauppinen$^{2}$ and P.~J.~Hakonen$^1$}
\affiliation{
$^1$Low~Temperature~Laboratory,~Helsinki~University~of~Technology,
Espoo, Finland\\
$^2$Center for New Materials,~Helsinki~University~of~Technology,~Espoo,~Finland\\
$^3$Chinese Academy of Sciences, Beijing, China}

\date{\today} 

\begin{abstract}
We have measured shot noise in single walled carbon nanotubes (SWNT)
with good contacts at 4.2 K at low frequencies ($f=600 - 850$ MHz).
We find a strong modulation of shot noise over the Fabry-Perot
pattern; in terms of differential Fano factor the variation ranges
over 0.4 - 1.2. The shot noise variation, in combination with
differential conductance, is analyzed using two (spin-degenerate)
modes with different, energy-dependent transmission coefficients. No
power law dependence of shot noise, as expected for Luttinger
liquids, was found in our measurements.

\end{abstract}
 \bigskip

\maketitle


Shot noise measurements have proven to be useful in providing information on the
fundamental conduction mechanisms in mesoscopic conductors \cite{BB}. For example, shot
noise has been utilized to determine the effective charge of quasiparticles in
fractional quantum Hall systems \cite{Saminad97,Picciotto97}. In multiterminal
conductors, current-current cross-correlations have been employed for investigating the
fermionic nature of charge carriers \cite{Henny99,Oliver99}. Also in single walled
carbon nanotubes (SWNT), noise is expected to be a valuable tool for studying the
physics of charged elementary excitations
\cite{Kane,Roche02,Pham,Trauzettel04,Kouwenhoven06,Yamamoto06,Kim06,Kontos}.

Liang et al. in Ref. \onlinecite{Park} have shown that SWNTs may act as molecular
waveguides for electronic transport. They employed a scattering matrix approach to show
that the results could be understood in terms of Fabry-Perot type of interference in
which reflection at the contacts played a crucial role \cite{Park,Peca03}. Shot noise in
the Fabry-Perot regime was recently studied at 4.2 K by Kim et al. \cite{Kim06} who
found power law dependence at low bias voltages as well as oscillations as a function of
bias voltage. These findings were assigned to Luttinger liquid behavior of SWNTs.

We have measured shot noise and AC conductance in a SWNT sample which displays a rather
asymmetric Fabry-Perot resonance pattern. The interference pattern has a strong
modulation, mostly dominated  by a single mode: the contribution from the second is only
about $0.1 \cdot 2 e^2/h$. We find a strong modulation of noise over the Fabry-Perot
pattern which we characterize in terms of a differential Fano-factor $F_d$. The
resonance condition is reflected as a strong suppression of shot noise with $F_d \simeq
0.4$ while the destructive interference yields $F_d \simeq 1.2$. We model our result
successfully using regular quantum conductor formalism with energy-dependent
transmission coefficients, without any recourse to Luttinger liquid (LL) physics.


The current in a quantum dot can be expressed in terms of energy
$\epsilon$ dependent transmission coefficient $ I = \frac{2e}{h}
\int_0^{eV} \sum_{i=1}^N \tau_i (\epsilon,V) d\epsilon$, where
$\tau_i (\epsilon,V)$ denotes the transmission coefficient of
spin-degenerate mode $i$ and we assume that the voltage is applied
to one terminal only. For differential conductance $G_d$, this
yields
\begin{equation}\label{dIdV}
    G_d=\frac{dI}{dV} = \frac{d}{dV} \left(\frac{2e}{h} \int_0^{eV}
    \sum_{i=1}^N \tau_i (\epsilon,V) d\epsilon \right).
\end{equation}
In the case of non-interacting electrons, $\tau(\epsilon,V)$ is voltage independent and
Eq. (\ref{dIdV}) reduces to $dI/dV= G_0 \sum_{i=1}^N \tau_i (eV)$ with $G_0 = 2e^2/h$.

The low-frequency shot noise is given by
\begin{eqnarray}\label{shot}
\b S(V) =\int dt\eavg{\delta I(t)\delta I(0)+\delta I(0)\delta I(t)} \\
\nonumber  = \frac{4e^2}{h} \int_0^{eV} \sum_{i=1}^N \tau_i
    (\epsilon)\left(1-\tau_i (\epsilon) \right) d\epsilon,
\end{eqnarray}
where $\delta I(t)= I(t)-\eavg{I(t)}$ at voltage $V$, and the current-current correlator
reduces to the latter form in the absence of interactions. By combining Eqs.
(\ref{dIdV}) and (\ref{shot}) we may define the differential Fano-factor as
\begin{equation}\label{diffF}
    F_d=  \frac{1}{2e} \frac{dS}{dV} / \frac{dI}{dV}=\frac{1}{2e} \frac{dS}{dI}.
\end{equation}
This is a quantity that we probe in our sensitive noise measurements
based on lock-in detection on modulated noise signal.

In the above formulas, it is assumed that $eV \gg k_B T$. In the cross-over regime with
$eV \sim k_B T$, one may write for the excess noise, the difference of current noise
and thermal noise
\begin{equation}\label{shotInter}
    S(I)-S(0) =  \frac{4 k_B T}{R(0)} \left( K F \frac{eV}{2k_B T}
    \coth \left(\frac{eV}{2k_B T}\right) -1 \right),
\end{equation}
where $S(0)$ specifies the noise at zero bias, $R(0)=V/I$ in the
limit $V \rightarrow 0$, $K=1$ and $F$ denotes the Fano-factor.
Formally, the left side can be identified as $\int_0^{I}
\frac{dS}{dI} dI= \int_0^{I} 2e F_d dI$. Hence, Eq.
(\ref{shotInter}) provides an interpolation formula for the
average Fano factor $\widetilde{F}= \frac{1}{I} \int_0^{I} F_d dI$
that is obtained from our measurements. Note that
$\widetilde{F}=(S(I)-S(0))/(2eI)$ is the quantity that is often used
to determine the Fano-factor \cite{Birk}; at large $V \gg k_B T/e$,
$\widetilde{F}$ is equivalent to $F$ in Eq. (\ref{shotInter}).
Finally, in order to take into account non-linearities of IV-curve
we set $K=R(0)/(I/V)$ in Eq. (\ref{shotInter}).

\begin{figure}

  \includegraphics[width=7.5cm]{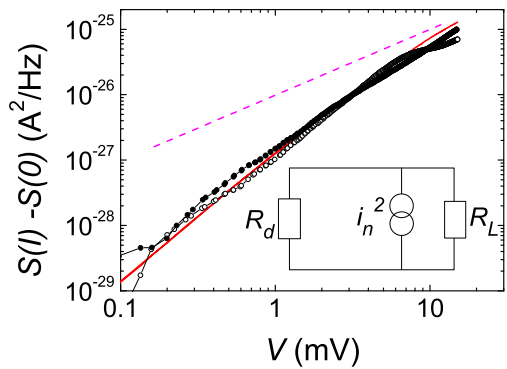}

  \caption{Excess noise $S(I)-S(0)$ as a function of bias voltage $V>0$ ($\circ$) and $V<0$ ($\bullet$).
  Red curve illustrates an evaluation of Eq. (\ref{shotInter}) using $F=0.65$ and the
  experimentally determined value $R(0)/(V/I)$. The dashed line
  refers to exponent $\beta = 1$.
  The inset displays the electrical equivalent model employed to calculate the coupling
  of the current fluctuations as well as the corrections due to non-linearities.} \label{ExpShot}

\end{figure}

In the nonlinear regime, noise measurements are sensitive to changes in the sample
resistance. For our setup, where a sample having a dynamic resistance of $R_d$ is
connected directly to a preamplifier with impedance $R_L=50$ $\Omega$, we may derive the
following equation (using the equivalent circuit displayed in the inset of Fig.
\ref{ExpShot})
\begin{equation}\label{FDexp}
    \frac{1}{G_{cal}} \frac{1}{R_L} \frac{\Delta P}{\Delta I}= 2e F_d - 2e F_d \frac{2 R_L}{R_d}
    - 2 i_n^2 R_d R_L \frac{\partial^2 I}{\partial V^2} ,
\end{equation}
which relates the measured, gain-adjusted noise power variation
$\frac{1}{G_{cal}}\frac{\Delta P}{\Delta I}$ to $F_d$ ($=  \frac{1}{2e} \frac{dS}{dV} /
\frac{dI}{dV}$). The second term on the right describes the first order correction in
measured shot noise due to changes in $R_d$ while the third term takes into account
corrections caused by the total system noise due to non-linearities, \textit{i.e.}
$i_n^2$ marks the full noise at the operating point, including the preamplifier noise.
The calibration constant $G_{cal}$ remains fixed within the factor $1+\frac{2R_L}{R_T} -
\frac{2R_L}{R_d}$, where $R_T$ denotes the resistance of the tunnel junction employed in
the calibration. In the data analysis, we neglect the corrections due to small changes
in $G_{cal}$ as well as the term $2e\frac{R_L}{R_d} F_d$ (as $\frac{R_L}{R_d} \ll 1)$,
but the non-linearity corrections are taken into account according to Eq. (\ref{FDexp})
by using the measured values for $\frac{\partial^2 I}{\partial V^2}$ and $S(I)$.


In our measurement setup, bias-tees are used to separate dc bias and the bias-dependent
noise signal at microwave frequencies. We use a liquid-helium-cooled low-noise amplifier
(LNA) \cite{Cryogenics04} with operating frequency range of $f=600 - 950$ MHz. The noise
signal is band limited to 600 - 850 MHz in order to avoid pick-up from mobile phones
working at 940 MHz despite of a Faraday cage. After amplification of 80 dB, the signal
was detected by a zero-bias Schottky diode with 0.5 mV/$\mu$W. A microwave switch and a
tunnel junction were used to calibrate the gain. For more details we refer to Ref.
\onlinecite{WuNoise}.

Noise was measured using three different methods (in the order of increasing
sensitivity): 1) noise at DC current 2) LOCK-IN detection of noise using sine-wave
modulation of current, $I=I_{DC}+\delta I \sin (\omega t)$ where $I_{DC} \gg \delta I$,
3) current modulation by square-wave $\Pi(t)$, $I=I_{DC}+I_m \Pi(t/t_0)$ where $I_{DC} =
I_m/2$. The calibration constants for each scheme were measured making similar
experiments on a tunnel junction sample of a resistance of $R_T=22$ k$\Omega$. The data
reported in this paper are mostly measured using method 2. Data obtained using method 3
agreed well with those measured with method 2, but the implementation of the
non-linearity corrections turned out to be more problematic for method 3 than for 2. In
addition to the above measurement schemes, we also performed experiments along method 2
where the DC bias was kept at zero and the noise modulation was measured at frequency $2
\omega$  while having the excitation at $ \omega$. By extrapolating the results of this
method to $\delta I=0$, we obtain the "zero-bias" Fano-factor. In the corrections using
Eq. (\ref{FDexp}), we have estimated $i_n^2 = 2.5\cdot 10^{-24}$ A$^2$ which is obtained
from the noise temperature $T_N=3.5$ K of our cooled LNA \cite{Cryogenics04} using
$i_n^2 = 4 k_B T_N/R_L$ for the unmatched case, as $4 k_B T_N/R_L$ is much larger than
the shot noise generator $S_i = \widetilde{F} \cdot 2eI$.


Our nanotube sample was made using surface CVD growth with Fe catalyst. The length was
$L=0.7$ $\mu$m and the diameter $\phi=2$ nm.
The contacts on the nanotube were made using standard e-beam overlay lithography. In the
contacts, 10 nm of Ti was employed as a sticking layer before depositing 70 nm of Al,
followed by 5 nm of Ti. The width of the two contacts was $200$ nm and the separation
between the them was 0.3 $\mu$m. The electrically conducting body of the silicon
substrate was employed as a back gate, separated from the sample by 100 nm of SiO$_2$.

\begin{figure}

 \includegraphics[width=7.5cm]{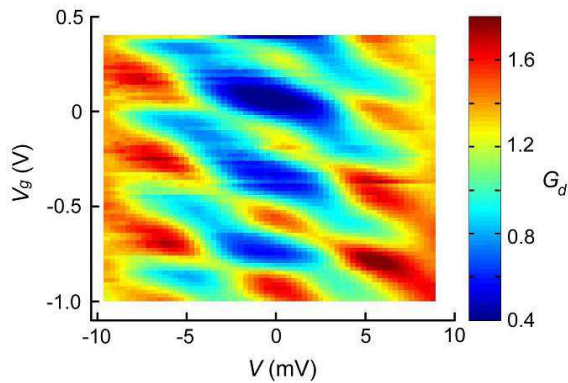}

    \caption{(Color online) Differential conductance $G_d$ on the plane spanned by
    bias voltage $V$ and gate voltage $V_g$. The scale bar is given on the
    right in units of $e^2/h$ ($=G_0/2$).} \label{SWNT-G}

\end{figure}

A scan of differential conductance $G_d=\frac{dI}{dV}$ is displayed in Fig.
\ref{SWNT-G}. Clear maxima/minima in $G_d$ with gate modulation are observed at zero
bias, but no characteristic features of odd/even effects that are found in the Kondo
regime of carbon nanotubes \cite{Linde00}. Therefore, we conclude that the pattern is
due to Fabry-Perot (FP) interference even though it appears more asymmetrically-striped
than observed typically \cite{Park,Kontos}. The maximum $G_d$ is only about $1.0 \cdot
G_0$ $ (=2e^2/h)$ which indicates a rather weak coupling to one of the orbital modes of
our nanotube sample (this will become more evident when discussing the shot noise data
below). The contrast of the fringes, more than 100\%, is clearly stronger than 10\% -
30\% found in Ref. \onlinecite{Park}. This may be connected with the fact that in our
sample we are dealing mostly with interference within one mode.

To reach the Fabry-Perot regime, the quality of contacts has to be good \cite{Park}.
This was investigated by making a separate cool down to 60 mK. No obvious change was
observed in the $G_d(V,V_g)$ pattern, indicating that "odd-even" Kondo features do not
appear even at dilution refrigerator temperatures. In this 2nd cool down, strong
proximity-induced supercurrent was observed in the nanotube, which is another indication
that the quality of the contacts is sufficient for the Fabry-Perot resonances.
Furthermore, in our third cool-down, we were able to observe Kondo-type features. Such a
change in the contacts is understandable as cool-downs are known to produce strain that
may alter the contact configuration by a tiny amount.

Shot noise data (excess noise $S(I)-S(0)$) using small voltage bias $V = 0.1 -10 $ mV at
$V_g=0.04$ V are displayed in Fig. \ref{ExpShot}. The data can be fitted using an
apparent power law $(S(I)-S(0)) \propto V^{\beta}$ with $\beta = 1.7$. This is clearly
larger than the exponent $\beta=0.64$ found in Ref. \onlinecite{Kim06} at $V = 0.1 -10 $
mV, also at 4.2 K. In Fig. \ref{ExpShot}, our data are compared with the cross-over
formula of Eq. (\ref{shotInter}) using a Fano-factor $F=0.65$, together with the
experimentally measured (voltage-dependent) ratio for K. The measured Fano-factor is not
exactly constant but, nevertheless, there is a good agreement between the measured data
and Eq. (\ref{shotInter}). Thus, we have to conclude that Luttinger liquid behavior is
not necessary to explain the power law dependence in our data.

\begin{figure}

 \includegraphics[width=7.5cm]{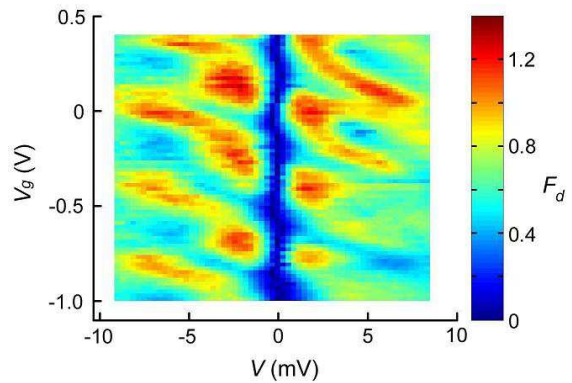}

    \caption{(color online) Differential Fano-factor $F_d$ on
    $V_g$ vs. $V$ plane. The scale bar is given on the right.} \label{SWNT-noise}

\end{figure}

The measured data on differential Fano-factor are displayed in Fig. \ref{SWNT-noise}.
The picture reflects, more or less, the pattern of $G_d$ in Fig. \ref{SWNT-G}: ridges of
large (small) $F_d$ follow the ridges of small (large) $G_d$ as would be expected for
non-interacting, one mode conductor. The swing of $F_d$, however, exceeds 1, which is
the upper limit in Landauer-Buttiker type of formalism for quantum dots and quantum
point contacts \cite{ADD}. At zero bias, $F_d$ appears to go to zero, which is an artefact due to
the employed AC-modulation scheme (method 2). At $V_g=0$, we checked the low bias Fano
by measuring noise at $2 \omega$ and varied the AC-modulation without any DC component.
We obtained $F=0.6 \pm 0.2$ as the modulation $\delta I \rightarrow 0$, which coincides
with a smooth continuation of the data in Fig. \ref{SWNT-noise}. Basically, the ridges
and gorges in Fig. \ref{SWNT-noise} continue, with small modulation, all the way down to
zero bias.

\begin{figure}

 \includegraphics[width=8cm]{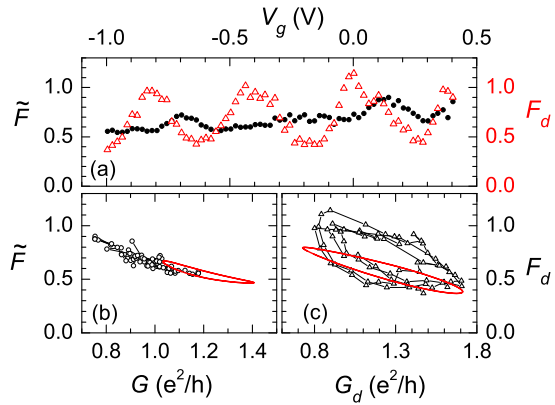}

  \caption{Plots obtained using data of
    Figs. \ref{SWNT-G} and \ref{SWNT-noise} at $V=-6.2$ mV.
    (a): Average differential Fano-factor $\widetilde{F}= \frac{1}{I} \int_0^{I} F_d dI $ ($\bullet$)
    and differential Fano-factor $F_d= \frac{1}{2e}
    \frac{dS}{dI}$ ($\bigtriangleup$) as a function of $V_g$.
    (b): $\widetilde{F}$ vs. total conductance $G=I/V$ plotted
    parametrically by varying $V_g$.
      (c): $F_d$ vs. $G_d$
    plotted parametrically by varying $V_g$. For the overlayed curves,
    see text. } \label{DiffPara1}

\end{figure}

In Fig. \ref{DiffPara1}, we interrelate the measured noise and conductance from Figs.
\ref{SWNT-G} and \ref{SWNT-noise} as suggested by Eqs. (\ref{dIdV}) and (\ref{diffF}).
At $V=-6.2$ mV in Fig. \ref{DiffPara1}a, $\widetilde{F}$ does not display any
oscillations as a function of gate voltage, only gradual large scale variation. The
relation between $\widetilde{F}$ and $G=I/V$, displayed in Fig. \ref{DiffPara1}b,
appears linear. At other bias values, this dependence is similar to that of the
differential quantities displayed in Fig. \ref{DiffPara1}c. The linear relation between
$\widetilde{F}$ and $G=I/V$ can qualitatively be explained using non-interacting
electron theory \cite{non-int}. For $F_d$, however, such an analysis does not work. The
dependence of $F_d$ and $G_d$ on $V_g$ is found oscillatory with a small relative phase
shift, which leads to ellipses in parametrically plotted curves of $F_d(V_g)$
\textit{vs}. $G_d(V_g)$ in Fig. \ref{DiffPara1}c. At $V>0$, the relations between noise
and conductance are similar, including unchanged rotation direction in the
parametrically plotted ellipses.

Clearly, in the presence of two modes and interactions, no unique solution can be
obtained for the transmission modes from the measurements of $G_d$ and $F_d$. We have
made a comparison with a phenomenological theory, assuming two modes with transmission
coefficients of the form
\begin{equation}\label{mode}
    \tau_i (\epsilon,V_g) = \overline{\tau}_{i} + m_i \cos \left[V_g/
    \Delta V_g \pm \epsilon/ (e\Delta
    V) + \varphi_i \right],
\end{equation}
where $i=1$ or 2 is the number of the mode, $\overline{\tau}_{i}$ denotes the average
value of their transmission, $m_i$ gives the modulation depth of $\tau_i$, and $
\varphi_1-\varphi_2$ specifies the relative phase difference of the transmission
modulation. This form is taken as the basis of the Fabry-Perot resonances, which produce
modulation of transmission coefficients along $V_g$ and $V$-axes with periods $\Delta
V_g$ and $\Delta V$, respectively. The curve in Fig. \ref{DiffPara1} illustrates the
result of a calculation for $\widetilde{F}$ and $F_d$ using Eqs. (\ref{shot}) and
(\ref{diffF}) with parameters: $\overline{\tau}_{1} = 0.48$, $\overline{\tau}_{2} =
0.13$, $ \varphi_1-\varphi_2=0$, and  $m_i=0.5 \overline{\tau}_{i}$ \cite{remark}.  The
model accounts for the main features of our data. It reproduces qualitatively deformed
ellipses in parametric plots of $F(V_g)$ and $F_d(V_g)$ \textit{vs}. $G_d(V_g)$, and the
elongated shape (even with zero width) tracks the variation found in the experiments.
According to the model, $\overline{\tau}_{1}$ dominates over $\overline{\tau}_{2}$ by a
factor of $3-4$. This agrees with the previous discussion that a description of our data
using a single dominant mode is a rather sound approximation.


In Ref. \onlinecite{BiercukPRL05}, lifting of mode degeneracies,
both spin and orbital, was observed as the conductance was quantized
in units of $e^2/h$. The asymmetry of the transport coefficient is
so large in our sample that there might almost be a lifting of
orbital degeneracy. However, since the asymmetry between
$\overline{\tau}_{1}$ and $\overline{\tau}_{2}$ became smaller in
subsequent cool-downs, we believe that the asymmetry comes rather
from the details of the contact configuration not from actual
removal of degeneracy. Thermally induced tension is a prime
candidate for such deformations \cite{Minot}. Moreover, there are
first principles calculations using realistic contact structures
which have yielded transport coefficients summing up to total
conductance around 1 $G_0$, both for Ti and Al contacts
\cite{PalaciosPRL03}, nicely in agreement with our conductance
results.


In summary, using conductance and shot noise measurements, we have obtained evidence for
quite asymmetric Fabry-Perot resonances in SWNTs. The Fano as well as the differential
Fano-factor, ranging between 0.4 - 0.9 and 0.4 - 1.2, respectively, were found to depend
on conductance either in linear or oscillatory fashion. We are able to explain our
findings using a phenomenological model with two (spin-degenerate) modes having
oscillatory, energy-dependent transmission coefficients of unequal magnitude. The large
observed values of $F_d$, however, point towards the importance of interaction effects,
which should be elaborated on theoretically in order to reach a full understanding of
our results.

We wish to acknowledge fruitful discussions with S. Andresen, M. Buttiker, G. Cuniberti,
R. Danneau, C. Glattli, F. Hekking, T. Heikkil\"a, T.~Kontos, L.~Lechner, B. Placais,
and P. Virtanen. This work was supported by the TULE programme of the Academy of Finland
and by the EU contract FP6-IST-021285-2.


\begin{thebibliography}{99}


\bibitem{BB} Ya. M. Blanter, M. B\"uttiker, Phys. Rep. \textbf{336}, 1
(2000).

\bibitem{Saminad97} L. Saminadayar, D. C. Glattli, Y. Jin and B.
Etienne, Phys. Rev. Lett. \textbf{79}, 2526 (1997).

\bibitem{Picciotto97} R. de-Picciotto, M. Reznikov, M. Heiblum, V. Umansky, G. Bunin,
and D. Mahalu, Nature (London) \textbf{389}, 162
(1997).

\bibitem{Henny99}  M. Henny, S. Oberholzer, C. Strunk, T. Heinzel,
K. Ensslin, M. Holland, and C. Sch\"onenberger,
 Science \textbf{284}, 296 (1999).

\bibitem{Oliver99}  W. D. Oliver, J. Kim, R.C. Liu, Y. Yamamoto, Science \textbf{284}, 299 (1999).

\bibitem{Kane} C. L. Kane and M. P. A. Fischer, Phys. Rev. Lett. \textbf{72}, 724
(1994).

\bibitem{Roche02} P. E. Roche, M. Kociak, S. Gueron, A. Kasumov, B. Reulet,
and H. Bouchiat, Eur. Phys. J. B \textbf{28}, 217 (2002).

\bibitem{Pham} K. V. Pham, F. Piechon, K. I. Imura, and P. Lederer,
Phys. Rev. B \textbf{68}, 205110 (2003).

\bibitem{Trauzettel04} B. Trauzettel, I. Safi, F. Dolcini,
and H. Grabert, Phys. Rev. Lett. \textbf{92}, 226405 (2004).


\bibitem{Kouwenhoven06} E. Onac, F. Balestro, B. Trauzettel,
C. F. J. Lodewijk, and L. P. Kouwenhoven, Phys. Rev. Lett.
\textbf{96}, 026803 (2006).

\bibitem{Yamamoto06} P. Recher, N. Y. Kim, Y. Yamamoto, cond-mat/0604613.

\bibitem{Kim06} N. Y. Kim, P. Recher, W. D. Oliver,
Y. Yamamoto, J. Kong, H. Dai, cond-mat/0610196.

\bibitem{Kontos} T. Kontos, \emph{et al.}, to be published.

\bibitem{Park} W. Liang, M. Bockrath, D. Bozovic, J. H.Hafner, M. Tinkham,
and H. Park, Nature, \textbf{411}, 665 (2001).

\bibitem{Peca03} C. S. Pe\c{c}a and L. Balents, K. J. Wiese, Phys. Rev. B \textbf{68},
205423 (2003).

\bibitem{Birk} H. Birk, M. J. M. de Jong, and C. Sch\"onenberger, Phys. Rev. Lett. \textbf{75}, 1610
(1995).

\bibitem{Cryogenics04} L. Roschier and P. Hakonen, Cryogenics \textbf{44}, 783 (2004).

\bibitem{WuNoise} F. Wu, L. Roschier, T. Tsuneta, M. Paalanen, T. Wang,
and P. Hakonen, Conference Proceedings of "24th International
Conference on Low Temperature Physics: LT24", Y. Takano, S. P.
Hershfield, S. O. Hill, P. J. Hirschfeld, and A. M. Goldman, eds.,
(AIP Conference Proceedings \textbf{850}, ISBN 0-7354-0347-3) pp
1482-1483.

\bibitem{Linde00} See, J. Nyg{\aa}rd, D. H. Cobden, and P. E. Lindelof,
Nature \textbf{408}, 342 (2000), and references therein.

\bibitem{ADD} But already taking in to account potential fluctuations due to charging
and decharging can give $F > 1$, at least when the IV-curve is hysteretic. See, e.g.,
Ya. M. Blanter and M. Buttiker, Phys. Rev. B \textbf{59}, 10217 (1999).

\bibitem{non-int} We have modelled the noninteracting transport as a sum of two modes with
transmission coefficients $\tau_1$ and $\tau_2$ and evaluated the Fano-factor by setting
$\tau_2=0.2$ and letting the other mode to account for the change in $G$: $\tau_1=
(G-\tau_2*G_0)/(G_0)$. The small contribution of the second mode, $\tau_2 \sim 0.2$, is
needed to lift $F(G)$ above the single mode curve $F=1-G/G_0$.


\bibitem{remark} The symmetry properties of $G(V,V_g)$ fix $ \varphi_1-\varphi_2 \sim 0$,
the average $G_d$ sets the sum $\overline{\tau}_{1} + \overline{\tau}_{2}$, while the
modulation of $G(V,V_g)$ ridges determines $\overline{\tau}_{1}/\overline{\tau}_{2}$.
The modulation depth of $\tau_i$ was taken the same for both modes, yielding the optimum
at 50\%.

\bibitem{BiercukPRL05} M. J. Biercuk, N. Mason, J. Martin, A. Yacoby, and C. M. Marcus,
Phys. Rev. Lett. \textbf{94}, 026801 (2005).

\bibitem{Minot} Strain alone cannot lift the degeneracy of the bands, see, \emph{e.g.},
E. D. Minot, Ph.D. thesis (Cornell University, 2004).

\bibitem{PalaciosPRL03} J. J. Palacios, A. J. Perez-Jimenez, E. Louis, E. SanFabian, and
J. A. Verges, Phys. Rev. Lett. \textbf{90}, 106801 (2003).


\end{thebibliography}
\end{document}